\documentclass[prl,aps,unsortedaddress,preprint]{revtex4}
\usepackage{epsfig}
\usepackage{dcolumn}
\usepackage{bm}

\usepackage{epsfig}
\usepackage[ansinew]{inputenc}

\begin{document}

\title{Experimental Observation of Electronic Coupling in GaAs Lateral Quantum Dot Molecules}

\author{L. Wang}
\email{l.wang@fkf.mpg.de}
\author{A. Rastelli}
\author{S. Kiravittaya}
\author{M. Benyoucef}
\author{O. G. Schmidt}
\affiliation{Max-Planck-Institut f\"ur Festk\"orperforschung, Heisenbergstr. 1, D-70569 Stuttgart, Germany}

\date{}

\begin{abstract}
We report the fabrication and photoluminescence properties of
laterally-coupled GaAs/AlGaAs quantum dots. The coupling in the
quantum dot molecules is tuned by an external electric field. An
intricate behavior, consisting of spectral line crossings and
avoided crossings is observed for different molecules. Anticrossing
patterns in the photoluminescence spectra provide direct evidence of
the lateral coupling between two nearby quantum dots. A simple
calculation suggests that the coupling is mediated by electron
tunneling, through which the states of direct and indirect exciton
are brought into resonance.
\end{abstract}

\pacs{73.21.La,78.55.Cr,78.67.-n}

\maketitle

Coupled semiconductor quantum dots (QDs) are attracting growing
interest due to their potential application as solid-state quantum
gates~\cite{loss98,burkard99}. Substantial progress towards the
experimental implementation of such quantum dot molecules (QDMs) has
been achieved in the last few years both for electrically defined
QDs (see Ref.~\cite{Koppens2006} and Refs. therein) and for
self-assembled, vertically-stacked
QDs~\cite{Krenner2005,Ortner05,Stinaff2006,Krenner2006}. In the
latter case, the coupling between two structurally different QDs is
controlled by applying a vertical electric field. A signature of
coupling and entanglement~\cite{Bester05} is represented by
anticrossing patterns in two-dimensional maps obtained from
photoluminescence (PL) spectra for different values of the external
field. To explore the possibility of coupling a larger number of
self-assembled QDs, investigations on lateral coupling are
needed~\cite{Bracker2006}. The fabrication of laterally-close QDs
with well-defined properties requires special growth
protocols~\cite{Schmidt2002,Songmuang2003c,Lippen04,Suraprapapich05,Hanke06,Yamagiwa06,Lee06,Beirne2006}
and, while indications of lateral coupling have been
reported~\cite{Unold05,Beirne2006}, anticrossing patterns for
lateral QDMs have not been observed so far.

In this Letter we employ a lateral electric field to tune the
coupling between two laterally-close GaAs/AlGaAs QDs and present the
observation of an anticrossing pattern in the PL spectra of a single
GaAs QDM.

The QDM samples are grown by molecular beam epitaxy (MBE) combined
with a method based on AsBr$_{3}$ selective etching of buried InAs
QDs and subsequent overgrowth. With a proper choice of etching and
overgrowth parameters, either single AlGaAs holes~\cite{Rastelli04}
or biholes aligned in the [110] direction can be created, which are
used as templates for the fabrication of either single QDs or QDMs.
To create QDMs, low density ($\lesssim 10^8$~cm$^{-2}$) InAs QDs are
first deposited at a substrate temperature of 500$^\circ$C on a GaAs
buffer, followed by a 30~s growth interruption. The substrate
temperature is lowered to 470$^\circ$C and 10~nm GaAs are deposited
while ramping the temperature back to 500$^\circ$C. An {\it in situ}
etching step with a nominal depth of 7.5~nm is then applied to
remove the buried QDs and obtain bow-tie shaped
nanoholes~\cite{Kiravittaya2003a,Songmuang2003c}, which are
overgrown with 10~nm Al$_{0.45}$Ga$_{0.55}$As. During
Al$_{0.45}$Ga$_{0.55}$As growth, single holes are found to split
into two closely-spaced holes aligned in the [110] direction. The
biholes are subsequently filled by depositing 1~nm GaAs followed by
a 1~min growth interruption. 100~nm Al$_{0.35}$Ga$_{0.65}$As, 20~nm
Al$_{0.45}$Ga$_{0.55}$As, and 10~nm GaAs complete the structure. The
GaAs-filled biholes embedded in AlGaAs represent QDMs below a thin
quantum well. Atomic force microscopy (AFM) in tapping mode is
employed for the morphological investigation of the bihole
structure. For this purpose, the sample is cooled to room
temperature immediately after the growth of the 10~nm
Al$_{0.45}$Ga$_{0.55}$As layer. Because of fluctuations inherent in
the self-assembled growth, the two QDs are generally not identical
and their mutual coupling can be controlled by an electric field
parallel to the [110] direction. To this end, interdigital gate
electrodes with 100/20~nm thick Au/Ti stripes and 30~$\mu$m spacing
are processed on the sample surface. Micro-PL ($\mu$-PL)
spectroscopy of single QDMs is performed by using a laser excitation
energy of 2.33~eV. The PL is analyzed by a 750~mm focal-length
spectrometer equipped with a Si charge-coupled device.

\begin{figure}[!th]
\epsfig{file=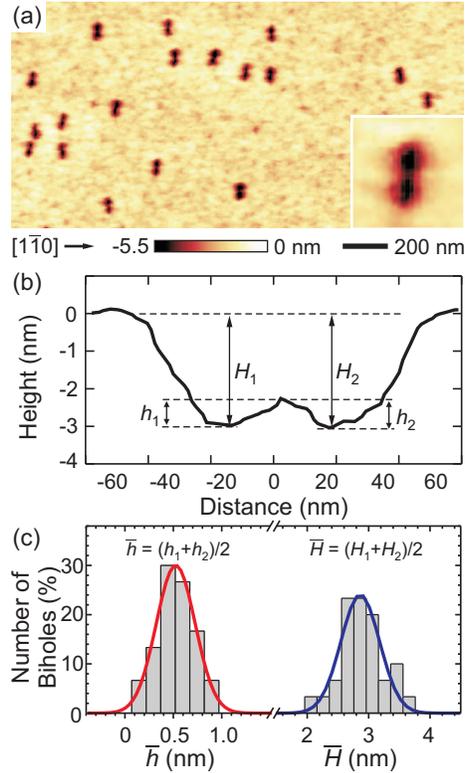} \caption{(a) AFM image of biholes on the
AlGaAs surface. The inset is a 125$\times$125 nm$^{2}$ zoom of a
bihole. (b) Linescan of a bihole along the [110] direction. (c)
Statistical analysis of hole depth and barrier height for biholes.}
\label{fig:1}
\end{figure}
Figure~\ref{fig:1}(a) shows an AFM image of biholes on the surface
of the 10~nm Al$_{0.45}$Ga$_{0.55}$As layer. The two holes are
normally slightly different in size and shape, but are invariably
aligned in the [110] direction. A typical linescan of a bihole along
the [110] direction is displayed in Fig.~\ref{fig:1}(b). The two
holes have an average center-to-center distance of 35$\pm$4~nm and
are separated by a thin barrier [see also the inset in
Fig.~\ref{fig:1}(a)]. Figure~\ref{fig:1}(c) shows a statistical
analysis of the hole depth and barrier height for the biholes.
Gaussian fits to the histograms show that the average depth and
barrier height of the biholes are 2.9~nm and 0.52~nm, respectively.
When the biholes are filled with GaAs and annealed for 1~min, the
deposited GaAs diffuses into the biholes, thus forming inverted
lateral GaAs QDMs aligned in the [110] direction.

\begin{figure}[h!]
\epsfig{file=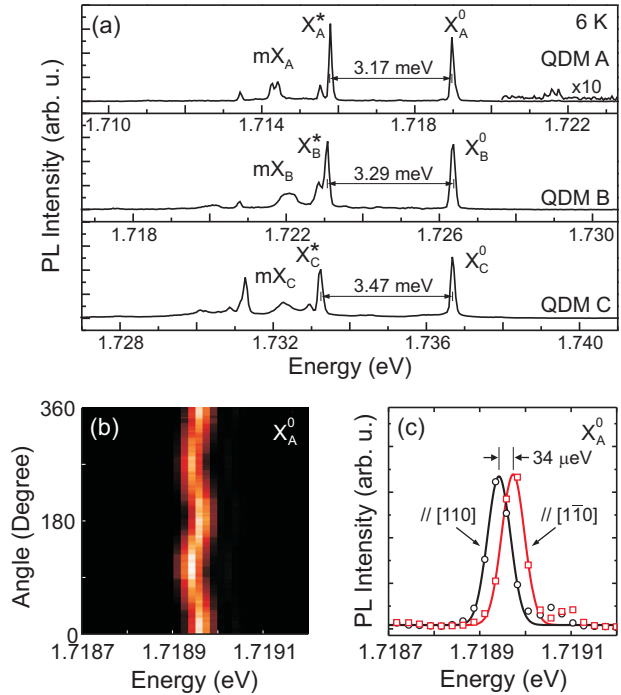} \caption{(a) $\mu$-PL spectra of three different QDMs. The energy axes are shifted to facilitate the comparison. (b)
Polarization dependent PL of the neutral exciton line in QDM~A. (c) PL spectra at the polarization angle of 0$^\circ$ and 90$^\circ$. The shift
of the peak position is due to the fine structure splitting of the neutral exciton line for QDM~A. Note that the spectra are fitted with
Lorentzian functions.}\label{fig:2}
\end{figure}

Typical PL spectra of three independent QDMs at relatively low
excitation power are shown in Fig.~\ref{fig:2}(a). Different QDMs
exhibit several common spectral features, which indicate that the
created QDMs have similar properties. The high-energy peaks, labeled
as X$^{0}$, are attributed to the recombination of an electron and a
hole confined in the larger QD composing a QDM (direct neutral
exciton). Another intense peaks X$^{*}$, well separated ($>$3~meV)
from the X$^{0}$ line, is present in all spectra. Since the
background doping in our samples is p-type, we tentatively assign
X$^{*}$ to a positive trion. Other features, generally labelled as
multi-excitons (mX) are also observed.

We now concentrate on the QDM~A [topmost spectrum in
Fig.~\ref{fig:2}(a)]. To confirm our identification of the spectral
lines, a polarization-dependent PL measurement is performed by
inserting a rotatable lambda-half waveplate and a fixed
Glan-Thompson polarizer in front of the spectrometer. The
oscillating peak energy as a function of polarization angle in
Fig.~\ref{fig:2}(b) indicates that the line X$^{0}$ originates from
a neutral exciton transition~\cite{Bayer02}. The two components of
the line X$^{0}_{A}$, corresponding to light polarized in the [110]
and [1$\bar{1}$0] directions, are plotted in Fig.~\ref{fig:2}(c).
The fine structure splitting deduced from the spectra is 34~$\mu$eV,
which is comparable to the values typically observed for ``natural''
GaAs QDs~\cite{Gammon1996}. Moreover, all other main lines do not
show any polarization dependence, which suggests that they originate
from charged excitons of direct or indirect nature~\cite{Bayer02}.

\begin{figure}[h!]
\epsfig{file=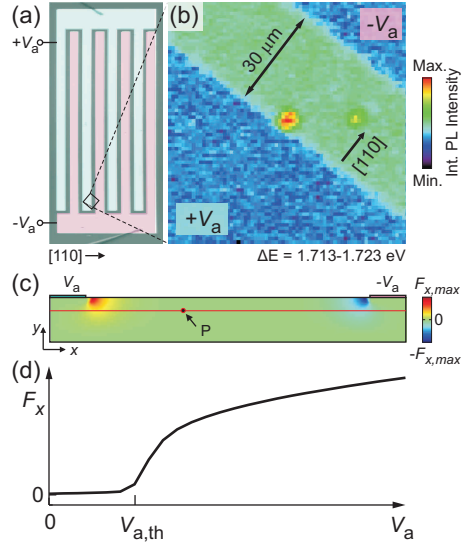} \caption{(a) Optical microscopy image of the interdigital Ti/Au gate electrodes on the sample surface. The height to
width ratio of the image is 2.5. (b) Integrated PL intensity map indicating the position of QDM~A close to an electrode. (c) Geometry of the
simplified structure used to estimate the lateral electric field $F_{X}$ as a function of applied voltage $V_{a}$. (d) Relation between $F_{X}$
and $V_{a}$ at the point P in (c).}\label{fig:3}
\end{figure}

Due to the slightly different size and shape of the two QDs in a
QDM, we expect that the X$^{0}_{A}$ emission originates from the
recombination of excitons in the larger dot. The recombination of
electron (hole) in the large dot with hole (electron) in the small
dot is suppressed, leading to a very weak signal in the PL spectra
[see the weak features at 1.721-1.722 eV in the QDM~A spectrum in
Fig.~\ref{fig:2}(a)]. However, when the QDM is subject to an
electric field, the emission of the QDM can be tuned through the
quantum confined Stark effect (QCSE), and therefore the interdot
coupling can also be tuned. The geometry of the Schottky
interdigital electrodes processed on the sample surface is shown in
Fig.~\ref{fig:3}(a). A voltage $\pm V_{a}$ applied to nearby
electrodes produces the desired electric field along the [110]
crystal direction.

In order to locate QDM~A with respect to the electrodes, we record
PL spectra while scanning the laser spot on the sample surface. (In
the experiment the sample is moved in a raster scan while the
excitation/collection optics remain fixed.) The signal is then
integrated over the spectral range of the QDM emission and a
corresponding PL intensity-map is displayed in Fig.~\ref{fig:3}(b).
From this map, we clearly see that the bright spot in the center of
the image, associated with the emission of QDM~A, is located about 2
$\mu$m away from one of the Ti/Au electrodes. Another spot (on the
right hand side of QDM~A) from another QDM is also visible.

While the contact structure employed here is easy to implement and
has  previously been used to apply a lateral field on QD
structures~\cite{Heller98,Beirne2006,Gerardot07}, the relation
between applied voltage $V_{a}$ and electric field is not trivial.
To clarify this point, the sample structure is simplified as a
two-dimensional geometry shown in Fig.~\ref{fig:3}(c). The GaAs
material parameters with a p-type doping of 10$^{15}$ cm$^{-3}$
(typical background doping level in our MBE chamber) are assumed
throughout the structure and the Schottky contacts are modeled as
heavily n-type doping regions (10$^{19}$ cm$^{-3}$). The Poisson
equation and the carriers' (electron and hole) diffusion equations
are simultaneously solved~\cite{fieldcal}. At $V_{a}$=0, the
built-in field along the $x$-direction $F_X$ in the depletion
regions can be clearly seen, as color-encoded in
Fig.~\ref{fig:3}(c). When the voltage is applied, the depletion
region around the left electrode extends laterally and the maximum
field strength increases. Figure~\ref{fig:3}(d) shows the behavior
of $F_{X}$ at a point P near the positively biased electrode
(1.5~$\mu$m from the electrode and 150~nm below the surface). The
result suggests that $F_{X}$ increases slowly until, at a certain
threshold value of the voltage ($V_{a,th}$), the depletion region
reaches P. At that point the field increases abruptly. Therefore we
expect that a pronounced QCSE will be observed only when
$V_a>V_{a,th}$, with $V_{a,th}$ depending on the distance between
QDM and the contact. This finding is in qualitative agreement with
the experimental observation that only QDMs close to one of the
contacts display a variation of the emission for $V_{a}$ less than
about 100~V.

\begin{figure}[!ht]
\epsfig{file=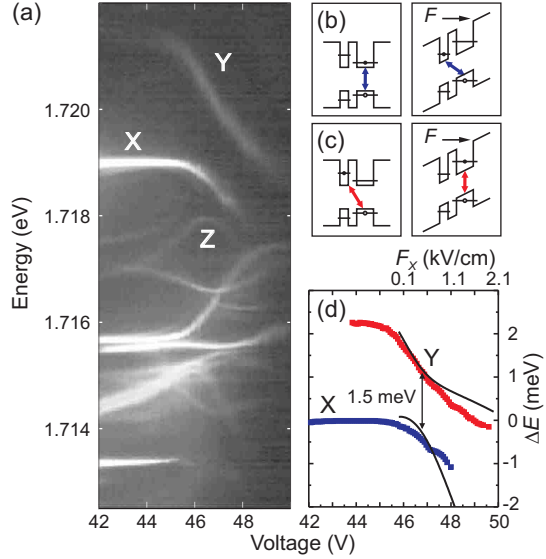} \caption{(a) Intensity map obtained from PL
spectra as a function of applied voltage for QDM~A. (b)-(c)
Schematic band diagrams with the relevant transitions at different
electric field amplitudes for X and Y lines, respectively. (d) Peak
positions of the X and Y lines observed in (a). Continuous lines
represent the result of the calculation discussed in the
text.}\label{fig:4}
\end{figure}
The first effect produced by moderate fields is a slight blue-shift
(of the order of 100~$\mu$eV) of most of the QDM lines. This
behavior, which we observed for all the investigated QDMs and also
for single GaAs/AlGaAs QDs, has not been reported for other
self-assembled QDs and is at present not understood. Most
importantly, for larger fields, the PL spectra of some of the
studied QDMs show pronounced red and blue-shifts leading to
intricate patterns consisting of crossings and avoided crossings
with increasing field. This is illustrated for QDM~A in
Fig.~\ref{fig:4}(a), which displays a PL intensity map obtained from
a series of spectra collected at 6~K as a function of applied
voltage (defined as 2$\times V_{a}$).

Since at present we are not able to identify unambiguously the
origin of the other spectral lines, we concentrate on the behavior
of the high energy side of the spectrum and in particular on the X
and Y lines. At low bias, X corresponds to the direct neutral
exciton X$^0$ transition. After the initial blue-shift, the onset of
a strong red-shift is observed at an applied voltage of about
44.5~V. The intensity of the X line then drops below the detection
limit. At low fields the spectral line Y is weak and can be assigned
to an indirect exciton recombination. When the voltage is increased
it gains in intensity and also red-shifts, but at a much larger rate
compared to X. Figure~\ref{fig:4}(d) shows the X and Y peak position
obtained by fitting the lines with Lorentzian functions. At the
beginning (for applied voltages below 46.8~V), the splitting between
these lines decreases. At a voltage of $\sim$46.8~V, the two lines
reach a minimum energy splitting of 1.5~meV and then gradually
separate. This anticrossing behavior, previously reported only for
vertically-stacked QDs~\cite{Krenner2005,Ortner05,Stinaff2006},
demonstrates that the two GaAs QDs composing our lateral QDMs are
quantum coupled. We can therefore interpret the energy splitting of
1.5~meV as the coupling energy. We also note that the pattern
deviates slightly from a perfect anticrossing. In particular, the X
line displays an anomalous shift at a voltage of about 48~V. This
anomaly is probably due to interaction of the energy levels
responsible for the X-Y transitions with other energetically close
states [see, in Fig.~\ref{fig:4}(a), the line Z, which approaches X
and avoids crossing it].

To obtain further insight into the anticrossing behavior, we perform
a quantized energy calculation of the electron and heavy hole
wavefunctions in a QD using a single band effective mass
approximation. In this calculation, a truncated pyramidal shape is
assumed for each GaAs QD in a QDM. First, the QD diameter is tuned
to fit with the $s$-shell and $p$-shell separation. Then, the height
of each QD in the QDM is adjusted to the ground state of the direct
and indirect transition of the QDM~A. The fitting of the
anticrossing energy is obtained by tuning the lateral distance
between the apex of each QD and varying the applied electric field
strength $F_{X}$. A splitting energy of 1.5~meV is obtained when the
separation distance is 31~nm, consistent with the value observed by
AFM (35$\pm$4 nm). Because of the large separation between the two
QDs, we expect the indirect exciton to have a large polarizability
and therefore display a pronounced QCSE at low fields. In the
calculation the anticrossing is in fact observed at a field
amplitude of only 0.45 kV/cm. The calculation also suggests that the
coupling is mediated by electron tunneling since hole tunneling
would yield smaller splitting energies. Schematic band diagrams of
the involved transitions at different field amplitudes for lines X
and Y are shown in Fig~\ref{fig:4}(b) and (c), respectively. By
assuming a linear relation between the electric field and the
voltage [$F_{X}=k\times(V_{a}-22.9~V)$, $k$=1000 cm$^{-1}$], we plot
the calculated energies as continuous lines in Fig.~\ref{fig:4}(d).
Since the model does not include all the states involved in the
transitions observed experimentally, it can not reproduce the
anomalous shifts. Moreover, the discrepancy between fit and
experiment at high fields is due to the nonlinear relation between
the field and the applied voltage [see Fig.~\ref{fig:3}(d)], a
technical issue which may be solved by improving the contact
structure as proposed in Ref.~\cite{Stavarache06}.

In conclusion, we have reported the fabrication of lateral GaAs QDMs
and we have provided evidence of lateral coupling between the two
nearby QDs. The quantum coupling is controllably tuned by applying
an in-plane electric field and manifests itself as an anticrossing
pattern in the PL spectra. The coupling is likely to be mediated by
electron tunneling, through which the states of direct and indirect
exciton are brought into resonance by the electric field. While the
present demonstration is based on a fully self-assembled structure,
we envision the possibility of using lithographically positioned
nanoholes~\cite{Schmidt2002,Kiravittaya06} as a template for the
fabrication of QDMs with well-defined position. The application of
an extra gate electrode above the barrier separating the two QDs may
allow the coupling strength to be tuned~\cite{Songmuang2003c}.

The authors thank M. Riek and T. Reindl for help in the sample processing and K. v. Klitzing for continuous support and interest. The work was
financially supported by the SFB/TR21, the BMBF (03N8711) and by the DFG research group ``Positioning of single nanostructures - Single quantum
devices''.



\begin{thebibliography}{27}
\bibitem[{fie()}]{fieldcal}
\bibinfo{note}{The finite element method with a commercial software package
  (COMSOL Multiphysics) is used for this calculation.}
\end{thebibliography}


\end{document}